\documentclass{aa}  

\usepackage{graphicx}
\usepackage{txfonts}
\usepackage{units} 
\usepackage[colorlinks=true, allcolors=blue]{hyperref}
\makeatletter 
\renewcommand*\aa@pageof{, page \thepage{} of \pageref*{LastPage}}
\makeatother

\begin{document} 

   \title{Constraining planetary mass-loss rates by simulating Parker wind profiles with Cloudy}

   \author{D.C. Linssen
          \inst{1}
          \thanks{E-mail: d.c.linssen@uva.nl}
          \and
          A. Oklop\v{c}i\'{c}
          \inst{1}
          \and
          M. MacLeod
          \inst{2}
          }

   \institute{Anton Pannekoek Institute for Astronomy, University of Amsterdam,
              Science Park 904, 1098 XH Amsterdam, The Netherlands
              \and
              Center for Astrophysics | Harvard \& Smithsonian, 60 Garden Street, MS-16, Cambridge, MA 02138, USA
             }

   \date{Received XX-XX-XXXX; accepted YY-YY-YYYY}

 
  \abstract
  {Models of exoplanet atmospheres based on Parker-wind density and velocity profiles are a common choice in fitting spectroscopic observations tracing planetary atmospheric escape. Inferring atmospheric properties using these models often results in a degeneracy between the temperature and the mass-loss rate, and thus provides weak constraints on either parameter. We present a framework that can partially resolve this degeneracy by placing more stringent constraints on the expected thermospheric temperature. We use the photoionization code \texttt{Cloudy} within an iterative scheme to compute the temperature structure of a grid of 1D Parker wind models, including the effects of radiative heating/cooling, as well as the hydrodynamic effects (expansion cooling and heat advection). We constrain the parameter space by identifying models that are not self-consistent through a comparison of the simulated temperature in the He 10830~Å line-forming region to the temperature assumed in creating the models. We demonstrate this procedure on models based on HD~209458~b. By investigating the Parker wind models with an assumed temperature between $4000$ and $12000$~K, and a mass-loss rate between $10^{8}$ and $10^{11}$~g~s$^{-1}$, we are able to rule out a large portion of this parameter space. Furthermore, we fit the models to previous observational data and combine both constraints to find a preferred thermospheric temperature of $T=8200^{+1200}_{-1100}$~K and a mass-loss rate of $\dot{M}=10^{9.84^{+0.24}_{-0.27}}$~g~s$^{-1}$ assuming a fixed atmospheric composition and no gas pressure confinement by the stellar wind. Using the same procedure, we constrain the temperatures and mass-loss rates of WASP-69~b, WASP-52~b, HAT-P-11~b, HAT-P-18~b and WASP-107~b.}

   \keywords{}

   \maketitle

\section{Introduction}
Atmospheric escape is thought to significantly influence the evolution of planets at short orbital distances, in particular for lower mass planets which can lose a sizeable fraction of their mass \citep{owen_atmospheric_2019}. It has been proposed that this process shapes planetary demographics, resulting in the observed hot Neptune desert and evaporation valley \citep{owen_kepler_2013, fulton_california-_2017, ginzburg_core-powered_2018}. However, planet formation \citep{lopez_how_2018, lee_creating_2022}, migration \citep{matsakos_origin_2016, owen_photoevaporation_2018} and the local environment \citep{kruijssen_bridging_2020} may play a role in sculpting these features in the population as well. Direct observational evidence of escaping atmospheres has come in the form of strong transit absorption in the Lyman-$\alpha$ line \citep[e.g.][]{vidal-madjar_extended_2003, lecavelier_des_etangs_evaporation_2010, lavie_long_2017}, the Balmer-$\alpha$ line \citep[e.g.][]{yan_extended_2018, yan_detection_2021} the helium line at 10830~Å \citep[e.g.][]{spake_helium_2018, allart_spectrally_2018, vissapragada_constraints_2020} and (ionized) metal lines in the UV \citep[e.g.][]{linsky_observations_2010, sing_hubble_2019, cubillos_near-ultraviolet_2020}.

Isothermal Parker wind models (\citealt{parker_dynamics_1958}, see also \citealt{lamers_introduction_1999}) are a common choice for modelling the upper atmospheric structure when fitting transmission spectra, such as those obtained in observations of the helium 10830~Å line. Parker wind models parametrize the outflow in terms of the temperature and mass-loss rate \citep{oklopcic_new_2018, lampon_modelling_2020, dos_santos_p-winds_2022}, enabling the retrieval of these parameters from observations, while staying agnostic about the mechanism(s) responsible for controlling the outflow.\footnote{An often-used complementary approach is based on self-consistent forward models, which assume the photoevaporation scenario of atmospheric escape to obtain a unique solution for a given set of parameters  \citep[such as][]{murray-clay_atmospheric_2009, salz_simulating_2016, wang_metastable_2021, caldiroli_irradiation-driven_2021}. } However, studies that have tried to constrain planetary mass-loss rates in this way have often found large degeneracies with temperature, resulting in only loose constraints on both parameters \citep[e.g.][]{mansfield_detection_2018, vissapragada_constraints_2020, palle_He_2020, lampon_modelling_2020, lampon_modelling_2021, paragas_metastable_2021}. In some cases, such a degeneracy can be somewhat reduced by fitting the full line profile observed at high resolution \citep{lampon_modelling_2021, dos_santos_p-winds_2022}. Since this is not possible for low-resolution observations such as those with the Hubble Space Telescope Wide Field Camera 3 \citep{spake_helium_2018, mansfield_detection_2018}, an ultra-narrowband photometric filter \citep{vissapragada_constraints_2020, paragas_metastable_2021}, or with the James Webb Space Telescope, a different approach to reducing the degeneracy is required for these cases. 

One promising way of reducing this degeneracy is observing multiple spectral lines for the same planet and performing a joint fit. For example, \citet{lampon_modelling_2020, lampon_modelling_2021} constrained mass-loss rates and H/He ratios by combining information from He~10830~Å and Ly$\alpha$ observations, but these analyses may be complicated by the fact that Ly$\alpha$ transits do not primarily probe mass-loss rates, as pointed out by \citet{owen_fundamentals_2021}. Interpreting these transmission signatures of the upper atmosphere will require detailed models that produce accurate excited and ionized level populations, taking into account non-local thermodynamic equilibrium (NLTE) processes \citep[e.g.][]{young_non-local_2020}. For the metal species absorbing in the UV, the atmospheric temperature plays an important role, as collisional effects may contribute significantly to their excitation and ionization state. With these goals in mind, we develop a framework around the NLTE plasma simulation code \texttt{Cloudy}\footnote{https://gitlab.nublado.org/cloudy/cloudy/-/wikis/home} \citep{ferland_cloudy_1998, ferland_2017_2017} to model the temperature structure of escaping atmospheres of exoplanets.

In this work, we use \texttt{Cloudy} to simulate the temperature structure of 1D Parker wind models. We thus combine the flexibility of parametrized density and velocity profiles with a detailed treatment of the chemical and thermal state of the outflow, instead of fixing an isothermal temperature. We show that the resulting temperatures can be used to identify Parker wind models that are not self-consistent. Restricting the `allowed' parameter space in this way allows for tighter observational constraints on the mass-loss rate, even when only a single spectral line is observed. We demonstrate our method here on a sample of planets with helium observations at 10830~Å, but it is not limited to this specific line. \citet{vissapragada_maximum_2022} recently pursued a similar goal using a different approach. They constrained the self-consistent Parker wind parameter space by analytically calculating the maximum mass-loss efficiency of an energy-limited photoevaporative outflow. 

The outline of this paper is as follows. In Sec. \ref{sec:parker_assumptions} we review the basic assumptions underlying isothermal Parker wind models and how we can test them to assess the model self-consistency. In Sec. \ref{sec:methods} we present an algorithm to model the temperature structure of upper exoplanet atmospheres with \texttt{Cloudy}. In Sec. \ref{sec:results} we present the resulting temperature structure constraints for different planets and combine these with fits to the observed line strengths to better constrain their mass-loss rates. In Sec. \ref{sec:discussion} we compare our results and methods to similar studies and discuss the advantages and potential drawbacks of our modeling approach. We summarize in Sec. \ref{sec:summary}.

\section{Assumptions of the Parker wind model} \label{sec:parker_assumptions}
The three equations that describe a steady-state planetary outflow in 1D are those of mass conservation (Eq. \ref{eq:mass_cons}), momentum conservation (Eq. \ref{eq:momentum_cons}) and energy conservation (Eq. \ref{eq:energy_cons}):
\begin{equation} \label{eq:mass_cons} 
\dot{M} = 4 \pi r^2 \rho v,
\end{equation}
\begin{equation} \label{eq:momentum_cons} 
v \frac{dv}{dr} + \frac{1}{\rho} \frac{dP}{dr} + \frac{GM_{p}}{r^2} = 0, 
\end{equation}
\begin{equation} \label{eq:energy_cons}
- \rho v \frac{d}{d r}\Bigg[ \frac{kT}{(\gamma - 1)\mu}\Bigg] + \frac{kTv}{\mu}\frac{d \rho}{dr} + \Gamma + \Lambda = 0. 
\end{equation} 
In these equations, $\dot{M}$ is the mass-loss rate, $r$ the distance from the center of the planet, $\rho$ the density, $v$ the outflow velocity, $P$ the gas pressure, $G$ the graviational constant, $M_p$ the planet mass, $M_*$ the stellar mass, $k$ the Boltzmann constant, $T$ the temperature, $\gamma$ the adiabatic index (5/3 for an ideal gas), $\mu$ the mean molecular weight and $\Gamma$ and $\Lambda$ the radiative volumetric heating and cooling rates, respectively. The gravitational influence of the star and the Coriolis force are neglected in Eq. \ref{eq:momentum_cons}. Solving this set of equations gives the three main parameters describing the outflow, $\rho(r)$, $v(r)$ and $T(r)$, but is a non-trivial task as it involves finding the radiative heating and cooling rates, which depend on $\rho$, $v$ and $T$. An example of a solution to this problem was provided by \citet{murray-clay_atmospheric_2009}, under the assumption that hydrogen photo-ionization and Lyman-$\alpha$ emission are the main contributors to $\Gamma$ and $\Lambda$, respectively. Recently, \citet{caldiroli_irradiation-driven_2021-1} presented a more advanced solution that includes additional contributors such as heating by helium ionization and cooling by bremsstrahlung and recombination.

Isothermal Parker wind profiles simplify the problem by ignoring Eq. \ref{eq:energy_cons}, taking $T/\mu$ to be constant, and solving only Eqs. \ref{eq:mass_cons} and \ref{eq:momentum_cons} for the density and velocity structure (see e.g. \citealt{lamers_introduction_1999} for the calculation). Maintaining a constant $T/\mu$ as $r \rightarrow \infty$ requires energy input as can be seen from the second term of Eq. \ref{eq:energy_cons}, which is not physical at very large altitudes. This is not a problem however at altitudes that are relevant for launching the outflow (until at least the Roche radius), since Parker wind structures are typically quite similar to self-consistent simulations such as those of \citet{salz_simulating_2016}, as shown in \citet{oklopcic_new_2018}. Even though the constant $T(r)/\mu(r)$ does not immediately imply an isothermal structure, a fixed temperature $T_0$ can be assigned if a fixed value for $\mu(r)$ is calculated. This $\bar{\mu}$ is usually taken as a weighted mean and requires assuming a gas composition \citep[see e.g.][]{oklopcic_new_2018, lampon_modelling_2020}. The simplification of the Parker wind model allows the freedom to choose $T_0$ and $\dot{M}$, which helps in retrieving these parameters for a given planet, as they can be adjusted to fit observational results. This is in contrast to self-consistent forward models that provide one outflow solution, and would have to alter their physical assumptions in order to match observations that are different from the model prediction \citep[e.g.][]{zhang_escaping_2022}.

In principle, a Parker wind solution can be calculated for any value of $T_0$ and $\dot{M}$ (as well as for different gas compositions), but not all of these models are physical, which can manifest itself in various ways. \citet{vissapragada_maximum_2022} presented one such assessment of the self-consistency of Parker wind profiles. Under the assumption of a maximally efficient energy-limited outflow heated by photoionization, they checked whether the total amount of heat absorbed by the atmosphere provided enough energy to launch the wind, which allowed them to rule out a part of the $T_0-\dot{M}$ parameter space for which this was not the case. 

Here, we assess the self-consistency of Parker wind solutions through a different route, focusing on the assumed temperature $T_0$. In particular, we investigate whether the assumed temperature of a model is reasonable, given its density and velocity structure and host star spectrum. We do so by reintroducing Eq. \ref{eq:energy_cons} and solving it for $T(r)$, using $\rho$ and $v$ of the Parker wind profile. The comparison between $T(r)$ and $T_0$ then acts as a `sanity check' on the Parker wind assumption that Eq. \ref{eq:energy_cons} is well approximated by the parametrized $T_0$. By quantifying the discrepancy between $T(r)$ and $T_0$, we construct a $T_0-\dot{M}$ Parker wind parameter space of self-consistent models. We note that a slightly more sophisticated approach would be to compare the parametrized $T/\mu$ to the $T(r)/\mu(r)$ structure we get from solving Eq. \ref{eq:energy_cons}. In literature however, Parker wind fits to observations are usually presented in terms of $T_0$ rather than $T/\mu$, and thus benefit more from constraints on $T_0$. Furthermore, for a few test models we found that the relative differences were similar for $T$ and $T/\mu$.

\section{Modelling exoplanet atmospheres with Cloudy} \label{sec:methods}
We used \texttt{Cloudy v17.02}, a NLTE plasma-simulation code suited for a wide variety of astrophysical conditions, to model Parker wind models. \texttt{Cloudy} assumes hydrostatic equilibrium and simulates a 1D slab of material under irradiation of a light source. In the context of our simulations, this means that given an atmospheric density profile and incident stellar spectrum, \texttt{Cloudy} calculates the (electron) temperature, ionization and excitation states of all of the most relevant atomic species, as well as the line and continuum radiative transfer and the associated heating and cooling rates, at each depth into the cloud. Alternative to letting \texttt{Cloudy} solve for the electron temperature, the code can also be run for a prescribed temperature profile, in which case it may return potentially unequal heating and cooling rates. The chemical composition of the plasma includes all elements up to atomic number 30 and can be changed as desired. We used the default solar composition throughout our simulations, but were forced to exclude the element calcium due to an error in \texttt{Cloudy}. For a more detailed description of \texttt{Cloudy}'s main features, we refer to \citet{ferland_2017_2017} or \citet{fossati_non-local_2021}, who provide a concise summary in their Section 2.2.1.

\subsection{Previous uses of Cloudy in exoplanet context} \label{sec:cluse}
When using \texttt{Cloudy} to simulate escaping exoplanet atmospheres, the assumption of hydrostatic equilibrium is not valid, and attention must be given to treating hydrodynamic effects properly. Perhaps the most notable example of this was provided by \citet{salz_simulating_2016}, who coupled \texttt{Cloudy} to magneto-hydrodynamics code \texttt{PLUTO} \citep{mignone_pluto_2012} to create \texttt{The PLUTO-CLOUDY Interface} \citep{salz_tpci_2015}. In their framework, \texttt{PLUTO} and \texttt{Cloudy} are alternately executed. \texttt{Cloudy} provides accurate net radiative heating or cooling rates, which are then passed on to \texttt{PLUTO} to calculate the next time step in the hydrodynamic simulation. By coupling to \texttt{PLUTO}, hydrodynamic effects on the temperature structure, such as heat advection and expansion cooling, are included. Using their code, \citet{salz_simulating_2016} simulated photoevaporative winds of 18 exoplanets in 1D. Their simulations are self-consistent and physically elaborate, but computationally more expensive than our approach, and therefore less suitable for exploring a large parameter space when fitting models to observations.

Alternatively to using \texttt{Cloudy}'s heating and cooling rates to solve for the temperature structure self-consistently, \citet{turner_detection_2020} and \citet{young_non-local_2020} used \texttt{Cloudy} with an assumed temperature-pressure profile to model the transmission spectra of KELT-9~b and HD~209458~b, respectively. They projected the one-dimensional T-P profile onto a spherically symmetric grid and ran \texttt{Cloudy} simulations for stellar light rays at different impact parameters from the center of the planet. Similarly, \citet{fossati_data-driven_2020} used this computational scheme with a family of T-P profiles to retrieve the atmospheric structure of KELT-9~b from published observations of the H$\alpha$ and H$\beta$ lines. \citet{fossati_non-local_2021} expanded on this work by forward modelling the T-P profile of KELT-9~b, using \texttt{Cloudy} to model the upper atmosphere ($P\lesssim10^{-4}$ bar) self-consistently. They ran \texttt{Cloudy} iteratively to solve for the temperature structure, which was then used to update the pressure scale heights and density structure. Their calculations were performed under the assumption of hydrostatic equilibrium, thus not considering an outflowing atmosphere nor determining mass-loss rates.

\subsection{Including hydrodynamic effects on the temperature structure} \label{sec:algorithm}
The temperature structure of the atmosphere is governed by the equation for energy conservation (Eq. \ref{eq:energy_cons}), which contains heating and cooling rates due to various physical processes. The first term of this equation represents advection: heating or cooling due to the bulk transport of the fluid. The second term represents cooling due to adiabatic expansion. The third and fourth term represent radiative heating and cooling, respectively, and are the only sources included in \texttt{Cloudy}. We developed an iterative algorithm centered around \texttt{Cloudy} in order to solve for the temperature structure while including advection and expansion heating and cooling rates. For a given density and velocity structure, the procedure is as follows. 

We start by running a \texttt{Cloudy} simulation at a constant temperature or previously solved temperature structure of a similar Parker wind profile (the initially assumed temperature does not affect the final temperature structure). In this fixed temperature mode, \texttt{Cloudy} returns the radiative heating and cooling rates as a function of altitude ($\Gamma$ and $\Lambda$ in Eq. \ref{eq:energy_cons}), as well as the ionization structure. We use the latter to calculate the mean molecular weight structure, considering the ionization fractions of hydrogen and helium and neglecting all other elements, which are included in the \texttt{Cloudy} simulations at solar metallicity. Together with the given density and velocity structures, we calculate the advection and expansion terms of Eq. \ref{eq:energy_cons}. 

Adding together all heating and cooling rates separately results in a ratio of total heating to cooling rate as a function of altitude, which we use to construct a new temperature structure. In regions where this ratio is $>1$, we construct a new temperature structure according to ${T_{new} = T_{old} (1 + 0.3 \mathrm{log}_{10}(\mathcal{H} / \mathcal{C}))}$, where $\mathcal{H}$ and $\mathcal{C}$ are the total heating and cooling rates, respectively. In regions where the ratio is $<1$, we construct a new temperature structure according to ${T_{new} = T_{old}/(1 - 0.3 \mathrm{log}_{10}(\mathcal{H}/\mathcal{C}))}$. We then pass the new temperature structure on to \texttt{Cloudy} to get $\Gamma$ and $\Lambda$ for the next iteration, and this process is repeated until we reach a converged temperature structure for which the total heating and cooling rate are in balance. The specific functional dependence of $T_{new}$ on the $\mathcal{H}/\mathcal{C}$ ratio does not have a physical motivation, rather, we found that this expression ensures relatively fast convergence of $T$. We set a convergence threshold of $\nicefrac{1}{1.1} < \mathcal{H}/\mathcal{C} < 1.1$, which for a typical temperature structure means $\lesssim 50$~K convergence. Higher precision is usually limited by round-off errors in \texttt{Cloudy}'s interpolation of the given density and temperature  structures onto its internal grid, combined with the fact that this internal grid changes throughout successive iterations of our algorithm, resulting in slight variations of the advection rate. Simulating a temperature structure in this way usually takes between 2 and 10 iterations.

In regions where advection is the dominant heating process, we are able to speed up the convergence. Instead of constructing a new temperature structure based on the $\mathcal{H}/\mathcal{C}$ ratio as described above, we construct the new temperature structure by solving Eq. \ref{eq:energy_cons} for $T$, using the $\mu$, $\Gamma$ and $\Lambda$ of the previous iteration. Since $\Gamma$ and $\Lambda$ depend on the temperature, this approach only works well when $\Gamma$ and $\Lambda$ of the newly constructed temperature structure are comparable to those of the last iteration, or if they are small compared to the advection and expansion rates. In an advection-dominated regime, the latter is indeed the case and we use this alternative way of solving for $T$, usually requiring on the order of $\sim$2 iterations to converge. In radiation-dominated regimes, it is exactly the dependence of $\Gamma$ and $\Lambda$ on $T$ that determines the temperature, and using $\Gamma$ and $\Lambda$ of the last iteration to construct the new temperature structure with Eq. \ref{eq:energy_cons} does not work, which is why we use the $\mathcal{H}/\mathcal{C}$ method described above.

\begin{figure}
   \centering
   \includegraphics[width=\hsize]{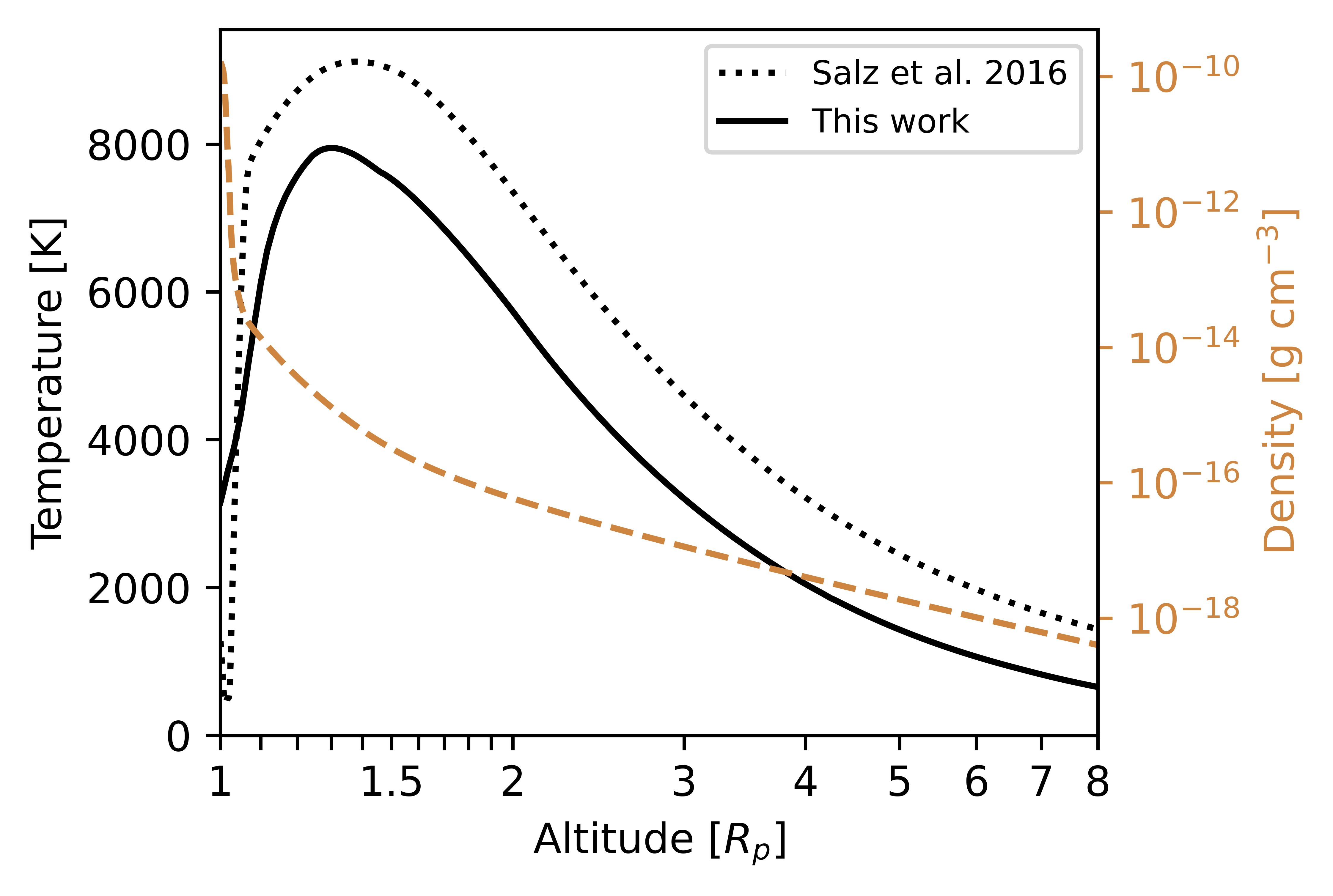}
      \caption{Comparison of simulated temperature structures for the density profile of HD~209458~b from \citet{salz_simulating_2016}. The orange dashed line shows the density structure (read from the right y-axis). Although not shown here, the qualitative behavior of the different heating and cooling rates as a function of altitude (such as shown for a different model in the bottom panel of Fig. \ref{fig:structure_bp209}) is similar to the results of \citet{salz_simulating_2016}, shown in their Fig. 5.}
         \label{fig:salzTcomp}
   \end{figure}

We tested our algorithm on the model of HD~209458~b provided by \citet{salz_simulating_2016}. They ran simulations with \texttt{The PLUTO-CLOUDY Interface} \citep[see also Sec. \ref{sec:cluse}]{salz_tpci_2015} of a photoevaporation driven escaping atmosphere. Their use of a hydrodynamics code ensures proper treatment of hydrodynamic thermal effects, and these simulations thus act as a useful validation of our code. We extracted the spectral energy distribution (SED) of HD~209458 from their Fig. 1 using the \texttt{WebPlotDigitizer}\footnote{https://automeris.io/WebPlotDigitizer/index.html} software package \citep{ankit_webplotdigizer_2020}. Using this SED in combination with the density and velocity structure shown in their Fig. 5, we converged on the temperature structure shown in Fig. \ref{fig:salzTcomp}\footnote{A reproduction package for all figures of this manuscript is available at doi:10.5281/zenodo.6798207}.

Our results agree quite well, but there is a constant $\sim1000$~K offset between the simulations. Part of this difference can be explained by \texttt{Cloudy}'s \texttt{wind advection} command, which was used by \citet{salz_simulating_2016}. This command includes the advection of ion species, such that neutral atoms are transported upwards with the outflow velocity, leading to higher ionization heating in the upper atmosphere. The command does not include the advection of heat (the first term of Eq. \ref{eq:energy_cons}), however. In our simulations, we did not use this functionality as it slows down \texttt{Cloudy} very significantly. We enabled it temporarily and observed an increase in temperature but were still unable to perfectly match the reported temperature structure of \citet{salz_simulating_2016}. Changing to the older \texttt{Cloudy} version {v13.01} that is used by \texttt{The PLUTO-CLOUDY Interface} also made a minimal difference. We could therefore not find the cause for the remaining discrepancy between the temperature structures.

\section{Results} \label{sec:results}
\subsection{Two types of temperature structure} \label{sec:twotypes}
\begin{figure}
   \centering
   \includegraphics[width=\hsize]{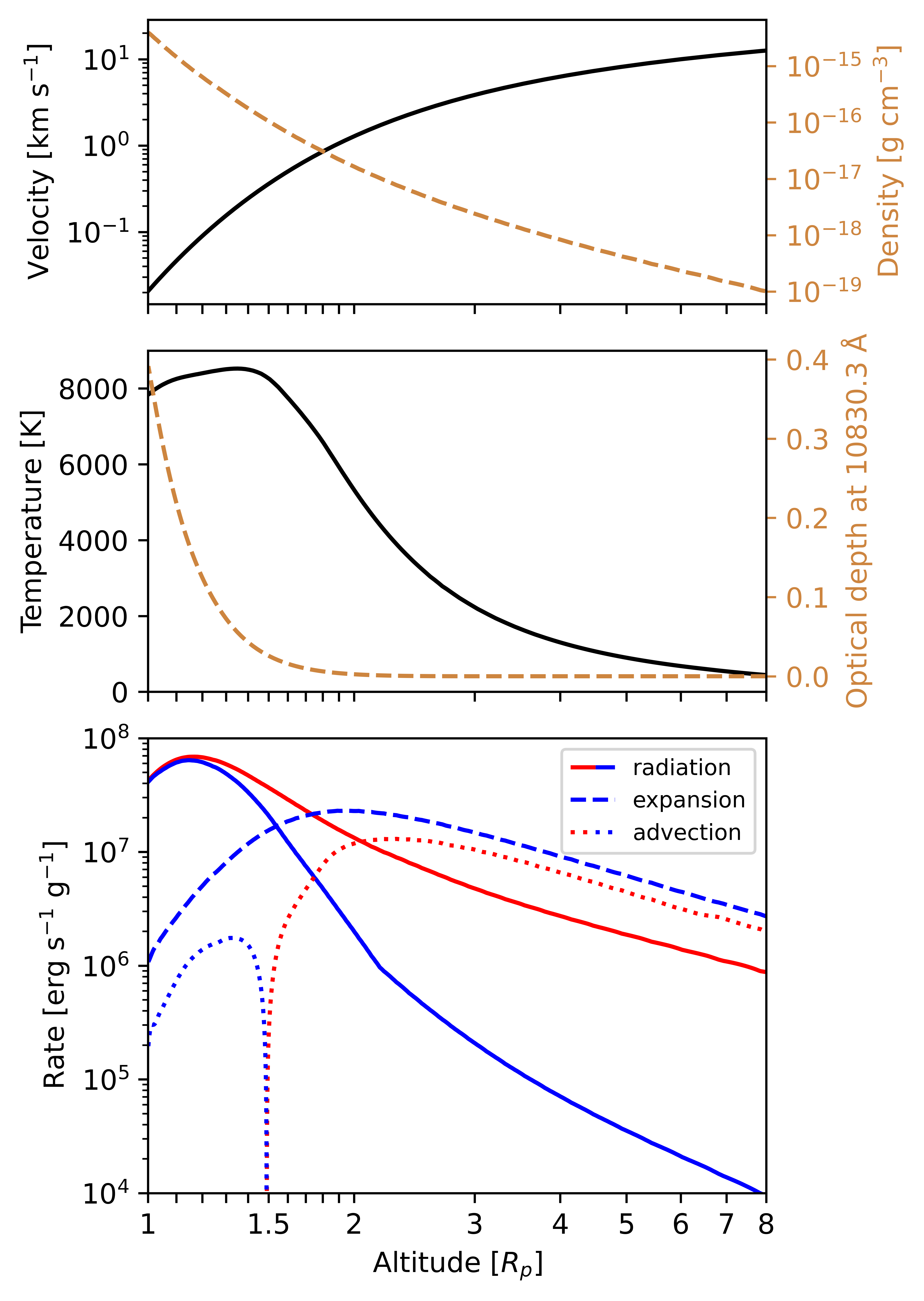}
      \caption{\textit{Top}: Velocity (black) and density (dashed orange, read from the right y-axis) structure for the Parker wind profile of HD~209458~b with $T_0=8000$~K, $\dot{M}=10^{10}$~g~s$^{-1}$. \textit{Middle}: Simulated temperature structure (black) and optical depth (dashed orange, read from the right y-axis) in the major component of the metastable helium line (dashed line). \textit{Bottom}: Relevant heating and cooling rates, shown in red and blue, respectively. We observe that at lower altitudes ($\lesssim2R_p$), radiative heating and cooling dominate, while at higher altitudes ($\gtrsim3R_p$), advection heating and expansion cooling are the dominant thermal processes.}
         \label{fig:structure_bp209}
   \end{figure}

We ran our code on a suite of Parker wind models for the well-studied exoplanet HD~209458~b, a hot Jupiter orbiting a G0 star. We used a gas composition of 90\% hydrogen and 10\% helium by number, a radius of $R_p=1.35R_J$ and a mass of $M_p=0.71M_J$ to create the models. We explored mass-loss rates between $10^8<\dot{M}<10^{11}$~g~s$^{-1}$ and temperatures between $4000<T_0<12000$~K. We used the SED of \citet{salz_simulating_2016}, shown in their Fig. 1, with a solar gas composition to run the \texttt{Cloudy} simulations. Fig. \ref{fig:structure_bp209} shows the result for the Parker wind profile with $T_0 = 8000$~K, $\dot{M}=10^{10}$~g~s$^{-1}$. The bottom panel shows the heating and cooling rates due to different physical processes. The qualitative behavior of these rates is the same across all calculated Parker wind models for this planet: radiative heating and cooling determine the temperature structure at lower altitudes, while advection heating and expansion cooling set the temperature at higher altitudes. Therefore, the temperature at lower altitudes does not depend sensitively on the outflow velocity.

\begin{figure}
   \centering
   \includegraphics[width=\hsize]{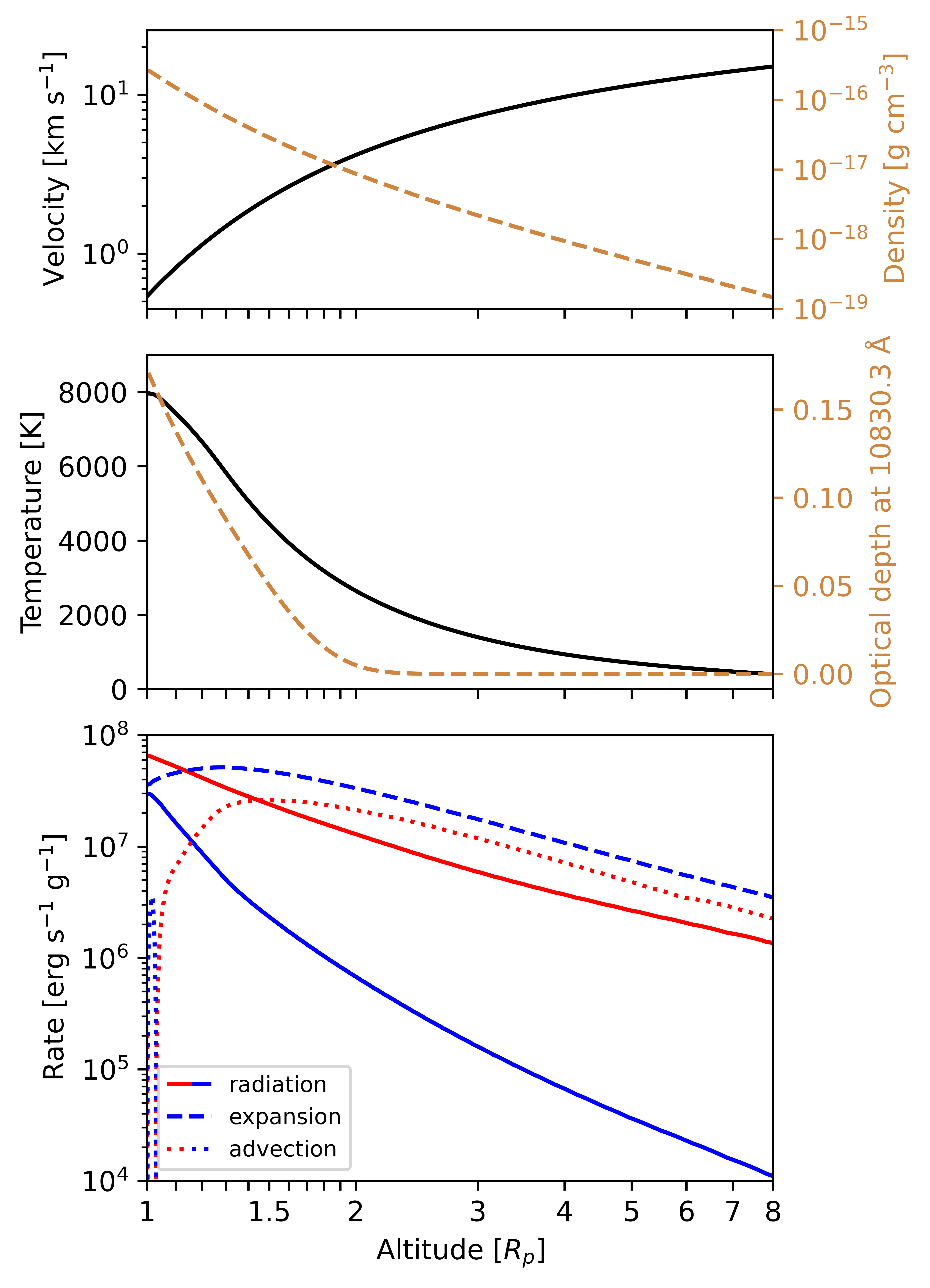}
      \caption{Similar to Fig. \ref{fig:structure_bp209}, but for the Parker wind profile of WASP-69~b with $T_0=6000$~K, $\dot{M}=10^{10}$~g~s$^{-1}$. Here, we observe that at nearly all altitudes, expansion cooling is the dominant cooling agent.}
         \label{fig:structure_bp69}
   \end{figure}

This behavior changes for planets with lower gravitational potential ($\vert \phi \vert= G M_p/R_p$). We ran a grid of Parker wind models for WASP-69~b, a Saturn-mass planet orbiting a K5 star. We adopted a mass of $M_p = 0.26M_J$, a radius of $R_p=1.057R_J$ and used the version 2.2 SED of the K6 star HD85512 from the MUSCLES survey \citep{france_muscles_2016, youngblood_muscles_2016, p_loyd_muscles_2016}. We ran models with mass-loss rates between $10^{9} < \dot{M} < 10^{11}$~g~s$^{-1}$ and temperatures between $5000 < T_0 < 9000$~K. Fig. \ref{fig:structure_bp69} shows a typical temperature structure, in this case for $T_0 = 6000$~K, $\dot{M}=10^{10}$~g~s$^{-1}$. WASP-69 b's lower gravitational potential compared to HD~209458~b (see Table \ref{tab:parsresults}) results in a qualitatively different atmospheric structure. For this type of planet, expansion cooling is the dominant coolant throughout the atmosphere while the temperature decreases monotonically with altitude for most models. A consequence of this is that the velocity influences the thermal structure even at the base of the atmosphere, as the expansion cooling rate depends on velocity (see Eq. \ref{eq:energy_cons}). 

When we turn to other planets (Table \ref{tab:parsresults} lists our sample, although we do not show their temperature structures), we find a similar behavior: planets with lower gravitational potential like HAT-P-18~b, HAT-P-11~b and WASP-107~b are dominated by expansion cooling throughout the atmosphere and showcase a monotonically decreasing temperature structure. WASP-52~b falls in the transition regime, with some Parker wind models dominated by expansion cooling and others by radiative cooling at lower altitudes. We thus identify two types of upper atmospheres. For high-gravity gaseous planets (loosely taken as $\vert \phi \vert \gtrsim 10^{12.82}$~erg~g$^{-1}$ here), radiative heating is balanced by radiative cooling and the thermospheric temperature increases with altitude until expansion cooling takes over and the exospheric temperature decreases with altitude. For lower gravity gaseous planets, expansion cooling always dominates over radiative cooling, resulting in generally cooler temperatures that decrease with altitude. A similar difference between the dominant cooling process in planets with high and low gravitational potential was identified by \citet{salz_simulating_2016} and \cite{caldiroli_irradiation-driven_2021-1}.

\subsection{Self-consistency of Parker wind profile temperatures} \label{sec:selfconparker}
For many models, the temperature structure we converged on with \texttt{Cloudy} does not agree with the constant temperature assumed to create the model. In principle, complete agreement is never expected, as the temperature structures found with \texttt{Cloudy} are not isothermal. Still, some models show reasonable overlap between the range of temperatures returned by \texttt{Cloudy} and the assumed constant value in the atmospheric regions that we are interested in. If this is not the case, such a discrepancy indicates that the Parker wind profile is not self-consistent and therefore not a physical model solution. While fitting observations, models that are not self-consistent can be excluded from consideration, which can help to put narrower constraints on the mass-loss rate and temperature. 

The criterion to assess and `quantify' the self-consistency of a Parker wind profile depends on the way in which \texttt{Cloudy}'s temperature structure is compared to the assumed isothermal profile. In principle, this choice is arbitrary and can be adjusted to fit the observations that we are aiming to interpret. In other words, one has the freedom to define a characteristic temperature of the converged structure that is suitable to the science case at hand, and compare it to the assumed temperature. 

In this work, we aim to aid the interpretation of metastable helium observations and we defined the characteristic temperature as the average in the He 10830~Å line-forming region, since it is this temperature that is probed by the observations. One way to calculate this (weighted) average temperature is to weigh the temperature in each bin by the contribution of that bin to the optical depth at 10830~Å. The problem with this definition is that higher altitude bins are relatively down-weighed, as their higher radial velocity shifts the line away from the rest-frame wavelength such that the optical depth contribution at 10830~Å (line center) is lower. To avoid this, we instead opted to weigh the temperature of each bin by that bin's metastable helium column density, which is independent of velocity but still directly related to the optical depth contribution. Thus using the metastable helium column densities $N_{He*}$ as given by our \texttt{Cloudy} simulations, we calculated the characteristic temperature
\begin{equation} \label{eq:weightedT}
    T_{He} = \frac{\sum T(r) \cdot N_{He*}(r)}{\sum N_{He*}(r)}.
\end{equation}
The difference between the temperature of the He 10830~Å line-forming region $T_{He}$ and the temperature assumed to create the Parker wind profile $T_0$ then provides a quantifier for the self-consistency of each Parker wind profile. Following this definition, the particular Parker wind profiles presented in Figs \ref{fig:structure_bp209} and \ref{fig:structure_bp69} are considered self-consistent, as $T_{He} \approx T_0$. As a measure for the spread of $T(r)$ around the characteristic temperature $T_{He}$, we calculated the weighted standard deviation
\begin{equation} \label{eq:weightedTsigma}
    \sigma_T^2 = \frac{\sum \big( T(r) - T_{He} \big) ^2 \cdot N_{He*}(r)}{\sum N_{He*}(r)}.
\end{equation}
This quantity indicates the range of atmospheric temperatures that would be probed by the metastable helium line, and thus can be used to judge the significance of the difference between $T_{He}$ and $T_0$.

\subsection{Mass-loss rate constraints} \label{sec:constraints}
\begin{figure*}
   \centering
   \includegraphics[width=\hsize]{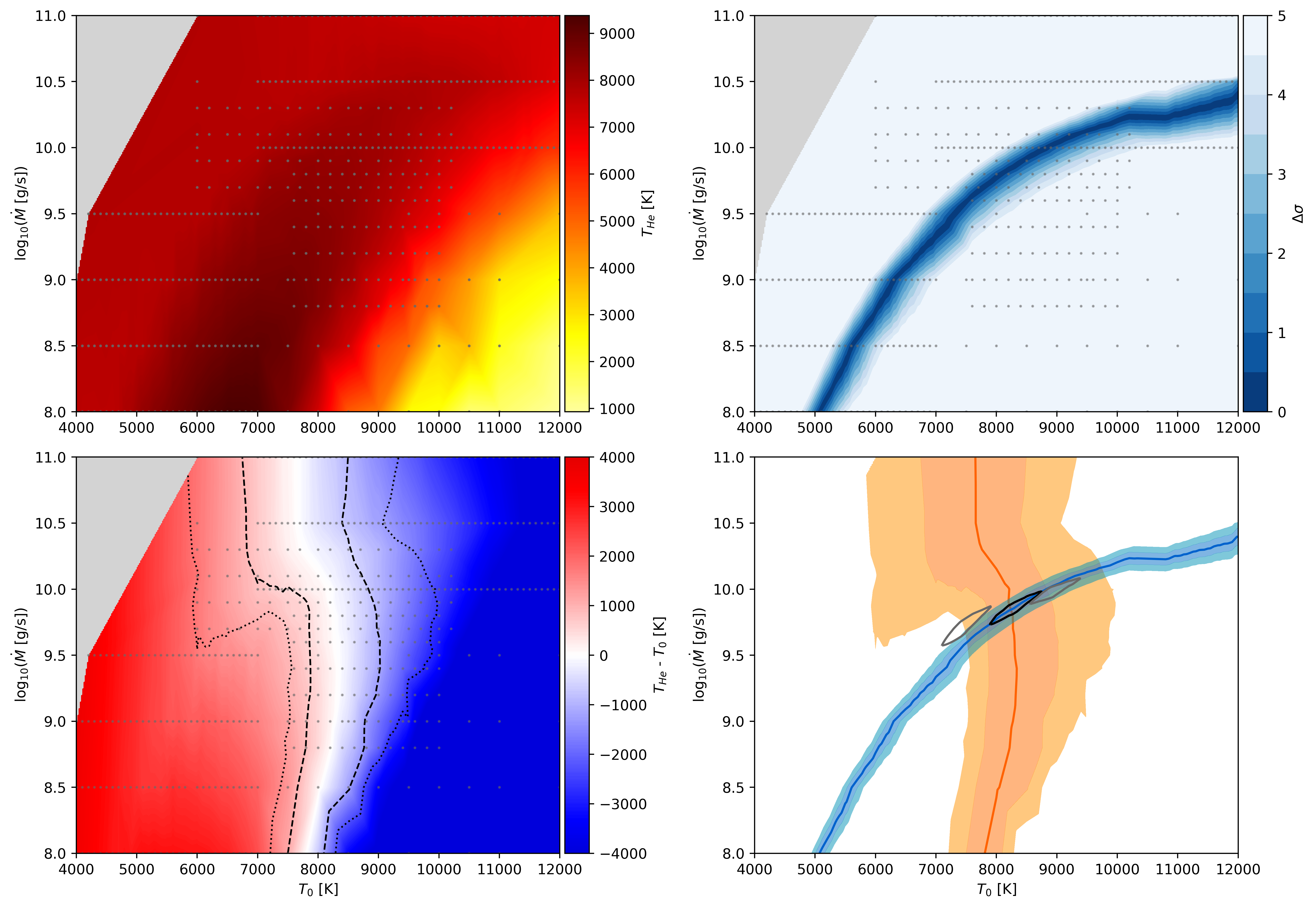}
      \caption{\textit{Top left}: helium line-forming temperatures $T_{He}$ (Eq. \ref{eq:weightedT}) for simulated temperature structures of HD~209458~b Parker wind models. The colormap linearly interpolates on the values of the models that were run (indicated by grey dots). The grey region in the upper left corner could not be simulated as \texttt{Cloudy} could not handle the high density of these models well. \textit{Bottom left}: Difference between $T_{He}$ (top left panel) and the assumed isothermal value $T_0$. The dashed and dotted lines indicate the $1\sigma$ and $2\sigma$ standard deviations of the self-consistent region based on Eq. \ref{eq:weightedTsigma}, respectively. \textit{Top right}: Compatibility between the He 10830~Å EWs of the Parker wind models and the observed value from \citet{alonso-floriano_he_2019}. The dark blue region indicates good agreement. \textit{Bottom right}: The orange line and dark and light areas are the self-consistent models, 1$\sigma$ and 2$\sigma$ contours from the bottom left panel, respectively, which we use as our prior. The blue line and dark and light areas are the best-fit models, 1$\sigma$ and 2$\sigma$ contours from the top right panel, respectively, which are our likelihood values. The 1$\sigma$ contour of the posterior is shown in black. The 1$\sigma$ posteriors of the 3x and 1/3x EUV flux simulations are shown in grey.}
         \label{fig:209_constraint}
   \end{figure*}

For each of our simulated Parker wind profiles for HD~209458~b, we calculated $T_{He} - T_0$ and $\sigma_T$. The results are shown in the left panels of Fig. \ref{fig:209_constraint}. We find a rather narrow range of self-consistent Parker wind profiles for which $T_{He} \approx T_0$, at an almost constant temperature of $T\sim8000$~K for the range of mass-loss rates explored. Our code always points us towards this temperature: models with $T_0 > 8000$~K result in helium line formation temperatures lower than $T_0$, as visible by the blue part in the bottom left panel of Fig. \ref{fig:209_constraint}, indicating that \texttt{Cloudy} prefers a cooler model. Vice versa, the red region in the figure for $T_0 < 8000$~K reveals that \texttt{Cloudy} prefers hotter models.

We next fit the Parker wind models to the He 10830 Å observations of HD 209458 b by \citet{alonso-floriano_he_2019}. They report an average absorption of 0.44\% in a 0.915~Å bandpass centered on the helium line. In order to estimate the uncertainty on this value, we extracted the light curve in this bandpass from their Fig. 6 using \texttt{WebPlotDigitizer} \citep{ankit_webplotdigizer_2020} and followed a bootstrap analysis similar to that described in Sec. 3.4 of \citet{salz_detection_2018}. We randomly drew half of the points in the light curve between second and third contact and resampled those points from normal distributions with a standard deviation of 0.167\%, which is approximately the average error on the light curve values. We calculated the mean of the new light curve and repeated this procedure 5000 times, resulting in a sample of average absorption values. The standard deviation of this sample was 0.0596\%, which we used as the error. Thus, the average absorption of $0.44 \pm 0.06$\% in a 0.915~Å bandpass could be translated to an equivalent width of EW=$4.03\pm0.55$~mÅ. 

For every simulated Parker wind profile, we calculate the transmission spectrum following the procedure described in Sec. 3.4 of \citet{oklopcic_new_2018}, using the metastable helium number density and temperature profile as given by the \texttt{Cloudy} simulations. We made mid-transit spectra at the planet's impact parameter while including stellar limb darkening using the parameters obtained from EXOFAST\footnote{https://astroutils.astronomy.osu.edu/exofast/limbdark.shtml} \citep{eastman_exofast_2013}. We integrated these helium transmission spectra to obtain the EW and compared it to the observed value. This approach means we did not fit the resolved spectrum because we want to develop a framework that works on both low- and high-resolution data. The fact that we fit the average in-transit absorption with a synthetic mid-transit spectrum can be justified in this case because of the relatively large scatter in the light curve, visible in Fig. 6 of \citet{alonso-floriano_he_2019}. Using the full in-transit light curve gives a higher S/N absorption value than using the few points around mid-transit. Furthermore, the in-transit light curve appears quite flat despite the scatter and does not clearly reveal asymmetric absorption indicative of a planetary tail or center-to-limb variations, which would otherwise prompt us to consider fitting only the mid-transit observations. The results of the fits are shown in the top right panel of Fig. \ref{fig:209_constraint}. A degeneracy between the mass-loss rate and temperature of good-fitting models is visible, which was also found by \citet{lampon_modelling_2020} who performed similar calculations. 

We then combined the observational and theoretical constraints into one joint constraint on the mass-loss rate and temperature. We do this in a Bayesian framework, treating the measure for self-consistency of the Parker wind model (bottom left panel of Fig. \ref{fig:209_constraint}) as our prior and the goodness-of-fit to the observations (top right panel of Fig. \ref{fig:209_constraint}) as the likelihood. For the prior, we convert the temperature difference $(T_{He}-T_0)$ to a p-value assuming it is normally distributed with standard deviation $\sigma_T$. For example, a temperature difference of $T_{He}-T_0 = 2 \sigma_T$ would be considered a 2$\sigma$ discrepancy and thus translate to a prior p-value of $\sim$0.05. For the likelihood, we convert the offset between modeled and observed EW to a p-value using the normally distributed observed error bar. Applying Bayes' theorem then gives the posterior distribution, which is shown in the bottom right panel of Fig. \ref{fig:209_constraint}. As visible in the figure, the constraints on self-consistent models complement the fits to the observations extremely well.

To estimate our modeling uncertainties, we repeated our calculations for host star spectra with different levels of EUV flux. The EUV part of the spectrum strongly influences the temperature because it is the main driver of radiative heating of the atmosphere through photoionization. Furthermore, the metastable helium number density depends on the EUV flux, since these photons populate the metastable helium state through ionization of ground-state atoms \citep{oklopcic_helium_2019}. Despite its importance, accurate stellar EUV spectra are notably hard to constrain through observations and thus not readily available. EUV spectra reconstructed through scaling relations (such as those used in this work) have typical uncertainties of one order-of-magnitude \citep{youngblood_EUV_2019, drake_pointing_2020}, which makes them a large source of uncertainty in our modeling efforts. In order to explore the effect of the EUV uncertainty on our constraints, we performed simulations throughout the $T_0-\dot{M}$ Parker wind parameter space with 3x and 1/3x flux at $\lambda < 1000$~Å. We followed the same steps as in the `fiducial' flux case to calculate the posterior distribution for both the high and low EUV flux case. The $1\sigma$ contours of these two posteriors are shown by grey dashed lines in the bottom right panel of Fig. \ref{fig:209_constraint}. In total then, we find a temperature of $T=8200 ^{+1200}_{-1100}$~K and a mass-loss rate of $\dot{M}=10^{9.84^{+0.24}_{-0.27}}$~g~s$^{-1}$, where we quote the fiducial simulations for our central value and the $1\sigma$ contours of the scaled EUV flux simulations for the upper and lower errors.

\begin{figure*}
   \includegraphics[width=\hsize]{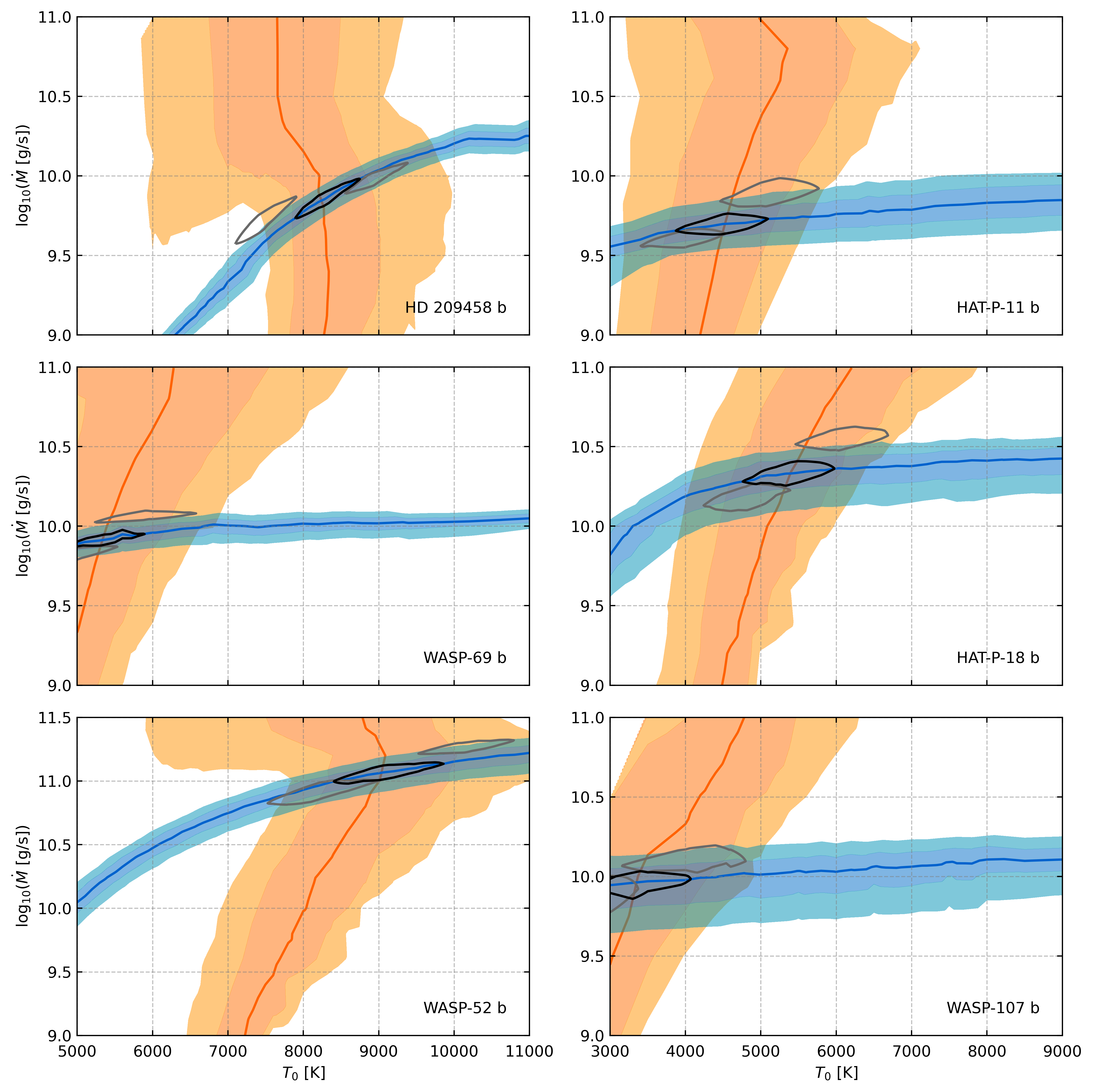}
      \caption{Similar to the bottom right panel of Fig. \ref{fig:209_constraint}, but also for a suite of other planets with reported metastable helium observations. The observations for HD~209458~b are by \citet{alonso-floriano_he_2019}, for WASP-69~b by \citet{vissapragada_upper_2022}, for WASP-52~b by \citet{kirk_kecknirspec_2022}, for HAT-P-11~b by \citet{mansfield_detection_2018}, for HAT-P-18~b by \citet{paragas_metastable_2021}, and for WASP-107~b by \citet{spake_helium_2018}. The constrained mass-loss rates and temperatures are listed in Table \ref{tab:parsresults}.}
         \label{fig:constrain_others}
   \end{figure*}

We repeated our analysis for five other planets. Table \ref{tab:parsresults} lists the input parameters, observed helium signal strengths and inferred temperature and mass-loss rate constraints for these planets, while Fig. \ref{fig:constrain_others} shows the posterior distributions. We chose our sample based on the availability of published metastable helium observations. The WASP-69~b observations were performed by \citet{vissapragada_constraints_2020} and reanalyzed by \citet{vissapragada_upper_2022}, who report an excess absorption of $0.512 \pm 0.049$\% in their narrowband filter. We convolved our helium model spectra with the transmission profile of this filter (S. Vissapragada, private communication) to obtain the excess absorption and compared these against the observed value. The WASP-52~b observations were performed by \citet{kirk_kecknirspec_2022}. We extracted the EW from their Fig. 9, which shows a mean absorption of $1.65 \pm 0.25$\% in a 2.44~Å bandpass, indicating an EW of $40.3 \pm 6.1$~mÅ. The HAT-P-11~b observations were performed by \citet{mansfield_detection_2018}. They report a white-light transit depth of $3371 \pm 15$~ppm and a $3560 \pm 34$~ppm transit depth in the 10809-10858~Å channel. Adding the uncertainties in quadrature, we obtain EW=$9.26 \pm 1.82$~mÅ. The HAT-P-18~b observations were performed by \citet{paragas_metastable_2021} and reanalyzed by \citet{vissapragada_upper_2022}, who report $0.70 \pm 0.16$\% excess absorption in the narrowband filter. The WASP-107~b observations were performed by \citet{spake_helium_2018}, who report EW=$48.02 \pm 10.78$~mÅ. We note that for WASP-69~b, HAT-P-11~b and WASP-107~b, additional spectrally resolved observations exist \citep{nortmann_ground-based_2018, allart_spectrally_2018, allart_high-resolution_2019, kirk_confirmation_2020, spake_posttransit_2021}, but they are generally consistent with the unresolved observations, so we took this opportunity to demonstrate the ability to constrain mass-loss rates from unresolved observations. With the exception of HD~209458, the SEDs of host stars in our sample are not available in the literature. Therefore, for each host star we used the closest spectral type SED from the MUSCLES survey \citep{france_muscles_2016, youngblood_muscles_2016, p_loyd_muscles_2016}, listed in Table \ref{tab:parsresults}. For HAT-P-11, which is a K4 star, we used an averaged MUSCLES spectrum of the K2 star HD~85512 and K6 star $\epsilon$ Eridani. We do not know how well all of these SEDs describe the true SEDs of our planets' host stars, especially in the EUV part of the spectrum.

\begin{table*}
\caption{Model parameters, helium line strengths extracted from observational papers, and the constrained temperatures and mass-loss rates for investigated exoplanets. The latter were inferred from Fig. \ref{fig:constrain_others}.} 
\label{tab:parsresults}
\centering
\begin{tabular}{l c c c c c c c c}
\hline\hline
Planet & $R$ & $M$ & $\mathrm{log_{10}}(\vert \phi\vert)$ & $a$ & SED & He $\lambda$10830 signal & $T$ & $\mathrm{log_{10}}(\dot{M}$)\\ 
& $[R_J]$ & $[M_J]$ & [cgs] & [AU] & & & [K] & [$\mathrm{log_{10}}$(g~s$^{-1}$)] \\
\hline                    
   HD~209458~b & 1.39 & 0.73 & 12.98 & 0.04747 & \citet{salz_simulating_2016} & $4.03 \pm 0.55$~mÅ \tablefootmark{1} & $8200^{+1200}_{-1100} $ & $9.84^{+0.24}_{-0.27}$\\
   WASP-69~b & 1.057 & 0.26 & 12.65 & 0.04525 & HD~85512 (K6)\tablefootmark{m} & $0.512 \pm 0.049$~\% \tablefootmark{2} & $5300^{+1300}_{-300\;u}$ & $9.91^{+0.19}_{-0.12\;u}$\\
   WASP-52~b & 1.27 & 0.46 & 12.82 & 0.0272 & $\epsilon$ Eridani (K2)\tablefootmark{m} & $40.3 \pm 6.1$~mÅ \tablefootmark{3} & $9000^{+1800}_{-1500}$ & $11.06^{+0.26}_{-0.25}$\\
   HAT-P-11~b & 0.389 & 0.0736 & 12.53 & 0.053 & Avg. K2 \& K6\tablefootmark{a} & $9.26 \pm 1.82$~mÅ \tablefootmark{4} & $4500^{+1300}_{-1100}$ & $9.70^{+0.29}_{-0.15}$\\
   HAT-P-18~b & 0.995 & 0.197 & 12.55 & 0.0559 & $\epsilon$ Eridani (K2)\tablefootmark{m} & $0.70 \pm 0.16$~\% \tablefootmark{5} & $5400^{+1300}_{-1200}$ & $10.33^{+0.29}_{-0.24}$\\
   WASP-107~b & 0.94  & 0.096 & 12.27 & 0.055 & HD~85512 (K6)\tablefootmark{m} & $48.02 \pm 10.78$~mÅ \tablefootmark{6} & $3400^{+1400}_{-400\;u}$ & $9.96^{+0.23}_{-0.19\;u}$\\

\hline
\end{tabular}
\tablefoot{
\tablefoottext{m}{From the MUSCLES survey, version 2.2 \citep{france_muscles_2016, youngblood_muscles_2016, p_loyd_muscles_2016}}
\tablefoottext{a}{Averaged MUSCLES spectra of $\epsilon$ Eridani (K2) and HD~85512 (K6)}
\tablefoottext{1}{EW from \citealt{alonso-floriano_he_2019}}
\tablefoottext{2}{Excess absorption from \citealt{vissapragada_upper_2022}}
\tablefoottext{3}{EW from \citealt{kirk_kecknirspec_2022}}
\tablefoottext{4}{EW from \citealt{mansfield_detection_2018}}
\tablefoottext{5}{Excess absorption from \citealt{vissapragada_upper_2022}}
\tablefoottext{6}{EW from \citealt{spake_helium_2018}}
\tablefoottext{u}{Underestimated uncertainty due to exceeding our modeled parameter space}}
\end{table*}

\section{Discussion} \label{sec:discussion}
\subsection{Comparison with other mass-loss rate estimates}
For HD~209458~b, we found a mass-loss rate of $\dot{M}=10^{9.84^{+0.24}_{-0.27}}$~g~s$^{-1}$ and a temperature of $T=8200^{+1200}_{-1100}$~K. This is a 2$\sigma$ difference with the mass-loss rate results of \citet{salz_simulating_2016}, who found $\dot{M} = 10^{10.27}$~g~s$^{-1}$ and a peak temperature of $T= 9100$~K for simulations of an photoevaporative wind. \citet{murray-clay_atmospheric_2009} employed a theoretical model of a hydrodynamically escaping atmosphere driven by photoevaporation and found a peak temperature of $T\sim8600$~K and a maximum mass-loss rate of $\dot{M} \leq 10^{10.5}$~g~s$^{-1}$, dependent on the UV flux of the host star. Both of these works estimated the mass-loss rate from simulations, while our constraint is based on observations in the metastable helium line. When comparing to other constraints on the mass-loss rate based on observations, we find that our estimates are slightly below the lower limit of $\dot{M} \geq 10^{10}$~g~s$^{-1}$ from \citet{vidal-madjar_extended_2003}, who observed HD~209458~b in the Ly$\alpha$ line. \citet{koskinen_characterizing_2010} later reanalyzed these and other observations in the O I triplet at 1304 Å with a parametrized atmospheric model, treating the thermospheric temperature as a free parameter. Their best-fitting models indicate a thermospheric temperature of $T=8000-11000$~K and a mass-loss rate of $\dot{M} = 10^{10-11}$~g~s$^{-1}$, showing reasonable agreement with our results.

For HAT-P-11~b, we found $\dot{M} = 10^{9.70^{+0.29}_{-0.15}}$~g~s$^{-1}$ and $T=4500^{+1300}_{-1100}$~K. This is a 2$\sigma$ difference with $\dot{M}=10^{10.29}$~g~s$^{-1}$ as reported by \citet{salz_simulating_2016}. It is also in disagreement with \citet{dos_santos_p-winds_2022}, who used the \texttt{p-winds} code to fit the high-resolution observations of the helium line by \citet{allart_spectrally_2018} with isothermal Parker wind models and found $\dot{M} = 10^{10.40\pm0.12}$~g~s$^{-1}$ and $T=7200\pm700$~K. Their temperature is 2$\sigma$ higher than ours, resulting in a higher mass-loss rate due to the degeneracy between these parameters. The discrepancy between the temperature inferred by \texttt{p-winds} and that predicted by \texttt{Cloudy} may be caused by inaccurate assumptions of either model. \texttt{Cloudy} might underestimate the temperature of HAT-P-11~b due to our adopted stellar spectrum (we use a template SED instead of the observed HAT-P-11 spectrum, which is not available) and/or atmospheric composition. On the other hand, \texttt{p-winds} might overestimate the temperature of HAT-P-11~b by fitting the line width with a model that does not include some line-broadening mechanisms which may be important, such as atmospheric circulation. These two codes use fundamentally different approaches to constrain the atmospheric temperature; hence, they are highly complementary. Understanding the cause of this discrepancy may ultimately shed new light on atmospheric properties and/or dynamics of HAT-P-11~b; however, this task is beyond the scope of this paper. 

For WASP-69~b, we found $\dot{M} = 10^{9.91^{+0.19}_{-0.12}}$~g~s$^{-1}$ and $T=5300^{+1300}_{-300}$~K, but the lower error bars are limited by the parameter space we ran and is in reality higher than 0.12~dex and 300~K (visible in the posteriors in Fig. \ref{fig:constrain_others}). The mass-loss rate we found is inconsistent with the value of $\dot{M}=10^{11.0}$~g~s$^{-1}$ that \citet{wang_metastable_2021} find while fitting the metastable helium line with their 3D hydrodynamic models. For WASP-107~b, our result of $\dot{M}=10^{9.96^{+0.23}_{-0.19}}$~g~s$^{-1}$ and $T=3400^{+1400}_{-400}$~K (with underestimated lower bounds, see Fig. \ref{fig:constrain_others}) is again not consistent with the result of \citet{wang_metastable_2021-1}, who find $\dot{M}=10^{11.3}$~g~s$^{-1}$ with their 3D hydrodynamic model fits to the helium line. This discrepancy is possibly caused by the fact that their outflow is always hotter than $T\gtrsim$7000~K and non-symmetric due to the inclusion of a stellar wind. For WASP-69~b, the model of \citet{wang_metastable_2021} has temperatures similar to our constrained value and the outflow geometry appears rather symmetric, so it is less clear where the discrepancy stems from. Finally, our self-consistent Parker wind parameter spaces for HAT-P-11~b, HAT-P-18~b and WASP-69~b are consistent with those identified in \citet{vissapragada_maximum_2022}.

For the observational He~10830~Å data of HD~209458~b, WASP-69~b, HAT-P-11~b and HAT-P-18~b, line fitting analyses using Parker wind profiles have already been published \citep{lampon_modelling_2020, vissapragada_constraints_2020, mansfield_detection_2018, paragas_metastable_2021}. In these works, the assumed constant temperature $T_0$ was used to calculate synthetic spectra. In Appendix \ref{sec:app}, we investigate the effect that using this temperature has on the resulting constraints and we compare our results to the literature. An interesting finding is that for the low-gravity planets, the blue curves of Fig. \ref{fig:constrain_others}, which show the fits to the observations, are much `flatter' than the blue curves of Fig. \ref{fig:constrain_others_f}. This means that using the non-constant temperature structures results in tighter constraints on the mass-loss rate compared to using the isothermal profiles.

\subsection{Comparison to other atmospheric escape modelling efforts}
The main advantage of our simulations lies in their applicability to interpreting observations. Parker wind profiles parametrize the planetary wind in two main parameters (assuming fixed chemical composition): temperature and mass-loss rate, allowing us to explore a range of possible outflow profiles without making assumptions on the physical processes driving the escape (e.g. photoevaporation versus core-powered mass-loss). This is supported by the relatively short run time of our code, roughly 15 CPU minutes to converge on the temperature structure of one Parker wind profile. 

Another advantage is that our simulations are NLTE through the use of \texttt{Cloudy}. The importance of NLTE effects was stressed by \citet{young_non-local_2020}, who found that some spectral lines were 40\% deeper than in the LTE case. In addition to this, \texttt{Cloudy} guarantees a detailed treatment of radiative processes. An example of one such process that is rarely included in simulations that treat radiative heating/cooling analytically rather than with a dedicated code such as \texttt{Cloudy}, is cooling by metal species. This was recently shown to be important in the upper atmosphere of KELT-9~b by \citet{fossati_non-local_2021}. They also used \texttt{Cloudy} to solve for the temperature structure but under the assumption of hydrostatic equilibrium, so our simulations differ by being applicable to outflowing atmospheres. Figs. \ref{fig:structure_bp209} and \ref{fig:structure_bp69} show hydrodynamic effects to be the dominant heating and cooling agents at high altitudes for both high- and low-gravity planets, and neglecting them would lead to an overestimation of the exospheric temperature. Finally, \texttt{Cloudy} supports the use of a full SED shape instead of representing the incident spectrum with a few flux bins, the latter being a common approach in atmospheric escape studies in literature \citep[e.g.][]{murray-clay_atmospheric_2009, shaikhislamov_3d_2020, caldiroli_irradiation-driven_2021}.

A potential drawback of our model is that we extrapolate the 1D sub-stellar solution to a 3D spherically symmetric structure to calculate the synthetic spectrum. However, the sub-stellar solution may not be representative of the part of the atmosphere that is probed with transmission spectroscopy, even if the outflow itself were spherically symmetric (which is most likely not the case).

\subsection{Using different characteristic temperatures}
In our Parker wind models, we compared the assumed temperature to the mean temperature in the helium line-forming region (Eq. \ref{eq:weightedT}), such that self-consistent profiles could be defined and found. This choice was specifically tailored towards the planets we investigated, as all of them have published observations in the He 10830~Å line, allowing for a meaningful comparison of the temperatures. The helium line is particularly suited for constraining the planetary outflow models, as the line-forming region is also the region where the bulk of the wind acceleration takes place. This can be seen by comparing the optical depth and velocity profiles in Figs. \ref{fig:structure_bp209} and \ref{fig:structure_bp69}. However, our approach is in not limited to planets that have been observed in the helium line. For planets with other observed atmospheric escape tracers, we could define a corresponding characteristic mean temperature weighed by the column density of the lower energy state. 

In some cases, it might even be desirable to compare the assumed temperature to some characteristic value of the simulated temperature structure that is not linked to any specific line. Examples are the temperature at some fixed altitude, or the peak temperature $T_p$. The latter ensures an unambiguous upper limit to the temperature of a planet, since $T_p < T_0$ implies that the whole atmosphere is inconsistent with the assumed temperature. We also tested this criterion for HD~209458~b and again found very similar temperature constraints to those of Fig. \ref{fig:209_constraint}. This implies that for this planet, the helium line forms in the hottest region of the atmosphere, which is indeed what we see in Fig. \ref{fig:structure_bp209}.

\subsection{Applicability to lower gravity planets}
For the class of planets with lower gravitational potential presented in Sec. \ref{sec:twotypes}, we lose some of the advantages that Parker wind profiles provide, as the dependence of the temperature on expansion cooling and, by extension, the outflow velocity at the base of the atmosphere exposes a stronger dependence on the mechanism by which the wind is launched. This is not the case for the class of higher gravity planets, where the resulting temperature structure does not depend on the velocity assumed at the base of the atmosphere. For the model of WASP-69~b shown in Fig. \ref{fig:structure_bp69}, the velocity at $1R_p$ is $v = 540$~m~s$^{-1}$, while the velocity of the model of HD~209458~b shown in Fig. \ref{fig:structure_bp209} is $v = 10$~m~s$^{-1}$ at $1R_p$. This illustrates that the Parker wind model of WASP-69~b does not transition into a hydrostatic atmosphere at low altitudes. Hydrodynamic simulations such as those of \citet{salz_simulating_2016} and \citet{caldiroli_irradiation-driven_2021} show a more physical picture, where even for low-gravity planets, the velocity approaches $v\sim 0$ at $\lesssim 1.1R_p$. Nevertheless, our method can provide parameter space constraints to aid in data interpretation because the simulated temperature structures should still hold for the given Parker wind profiles. However, because in the low gravitational potential regime Parker winds do not correctly model the lower altitudes, the contribution of expansion cooling may be too strong. Therefore, especially for spectral lines originating at lower altitudes, the temperature we constrain is potentially too low, in turn also resulting in lower mass-loss rate estimates.

\subsection{Energy-limited mass-loss efficiencies}
An often-used way of estimating planetary mass-loss rates is to assume that some given fraction of the incoming high-energy flux is converted into expansion work that drives the atmospheric mass loss. This mechanism has been dubbed `energy-limited' escape and dictates that the mass-loss rate depends linearly on the incoming XUV flux \citep[e.g.][]{owen_atmospheric_2019}:
\begin{equation} \label{eq:energylimited}
    \dot{M} = \eta\frac{\pi R_p^3 F_{XUV}}{G M_p K},
\end{equation}
where $\eta$ is the heating efficiency and
\begin{equation}
    K=1-\frac{3}{2}\frac{R_p}{R_{Roche}} + \frac{1}{2}\Bigg( \frac{R_p}{R_{Roche}}\Bigg) ^3
\end{equation}
a correction term for the fact that matter need only be lifted to the planetary Roche lobe to escape \citep{erkaev_roche_2007}. There exist many variations of Eq. \ref{eq:energylimited}, mainly due to the fact that the planet absorbs high-energy radiation over an area $\pi R_{XUV}^2 > \pi R_p^2$ that should then be averaged over the full planet surface area, which some authors include explicitly in the energy-limited formula. We choose to instead absorb these effects into $\eta$. In principle any planet undergoing atmospheric mass loss will satisfy Eq. \ref{eq:energylimited} for some value of $\eta$, but an `energy-limited' outflow usually refers to the case where $\eta$ is of order unity.

The energy-limited prescription finds its use mainly in population studies of escape, when mass-loss rates need to be estimated for a large sample of planets and performing individual hydrodynamical simulations is too computationally expensive \citep[e.g.][]{lammer_determining_2009, owen_evaporation_2017, rogers_exoplanet_2021}. A main difficulty for such analyses lies in finding the value of $\eta$ and how it depends on planetary parameters. Many works have addressed this problem through different avenues \citep{owen_planetary_2012, shematovich_heating_2014, owen_uv_2016, caldiroli_irradiation-driven_2021-1}. Although this manuscript is not a systematic investigation into the value of $\eta$, we still find it illustrative to plug the mass-loss rates we constrained for our sample of 6 planets into Eq. \ref{eq:energylimited} and compare the corresponding heating efficiencies to the literature. This comparison is especially useful because our mass-loss rates are constrained from observations, while most literature studies use theoretical predictions to evaluate $\eta$. 

   \begin{figure}
   \includegraphics[width=\hsize]{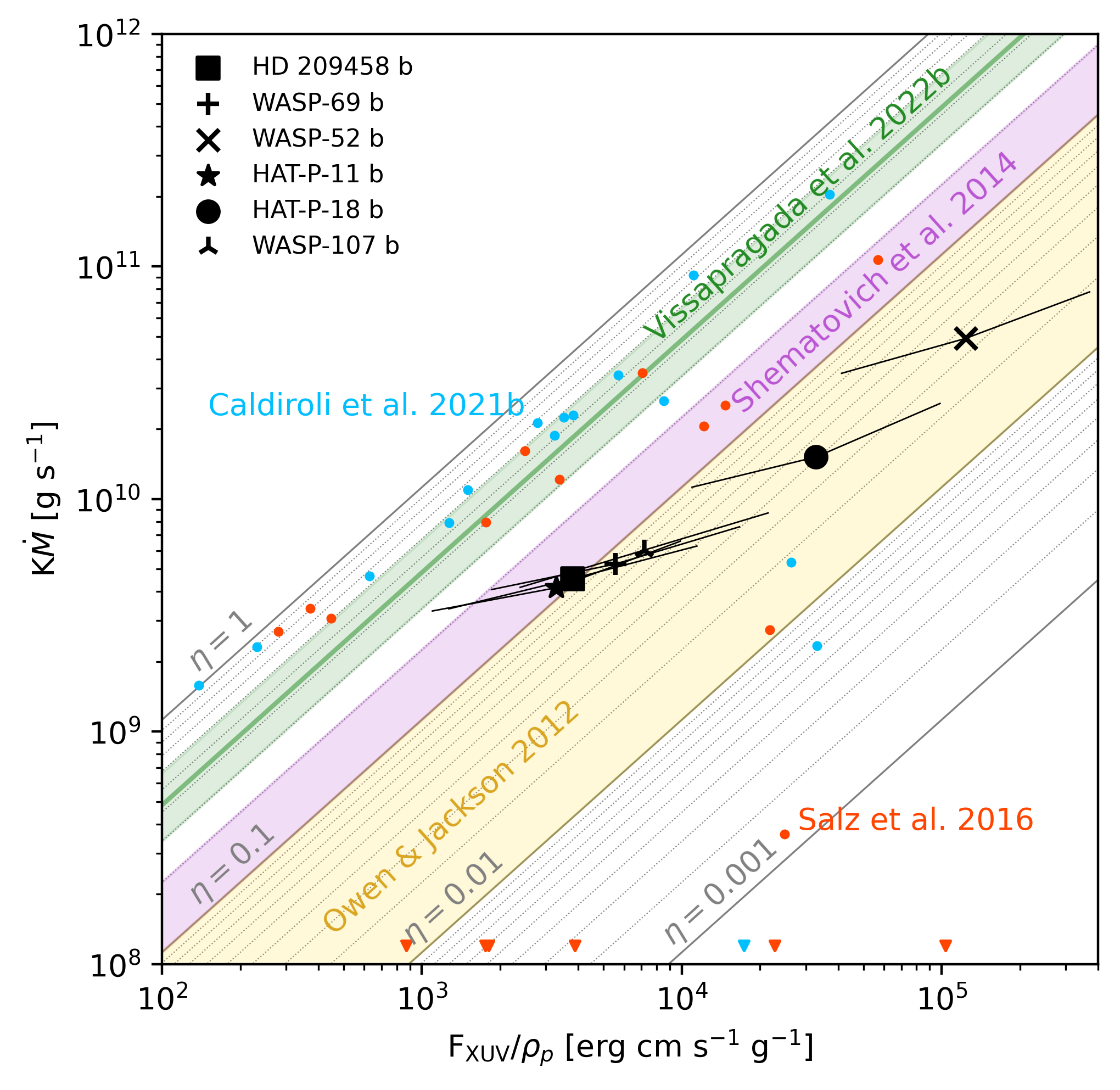}
      \caption{The black scatter points show our constrained mass-loss rates corrected by $K$ as a function of the ratio of high-energy irradiation to bulk density. The error bars indicate the results for the simulations at 1/3x and 3x EUV flux. The orange and blue scatter points show the results for the hydrodynamic simulations of \citet{salz_simulating_2016} and \citet{caldiroli_irradiation-driven_2021-1}, respectively. The yellow and purple shaded regions indicate the constraints on the efficiency parameter of \citet{owen_planetary_2012} and \citet{shematovich_heating_2014}, respectively. The green line and shaded region mark the best-fit and 1$\sigma$ contour of the efficiency constraint from \citet{vissapragada_upper_2022}.}
         \label{fig:energylimited}
   \end{figure}

Inspired by Fig. 4 of \citet{caldiroli_irradiation-driven_2021-1}, in Fig. \ref{fig:energylimited} we show $\dot{M}/K$ versus $F_{XUV}/\rho_p$ of our sample together with various estimates from the literature, where the choice of axes ensures $\eta$ is read off easily. Including the EUV flux uncertainty, our sample spans efficiencies of roughly 2\%-30\%. This range generally agrees with the results of \citet{shematovich_heating_2014}, who calculate a heating efficiency of 10\%-20\% for hydrogen-dominated upper atmospheres similar to HD~209458~b. \citet{owen_planetary_2012} explored X-ray driven outflows of different planetary masses and radii and found efficiencies in the range 1\%-10\% for gas giant planets. We marked this region of $\eta$-space completely in Fig. \ref{fig:energylimited}, but note that in principle their efficiencies depend on the planetary parameters. Furthermore, their definition of the efficiency is different from Eq. \ref{eq:energylimited} up to a factor of a few. Despite these considerations, the reported efficiencies seem generally consistent with our results. However, we do find somewhat lower efficiencies than the hydrodynamic simulations of \citet{salz_simulating_2016} and \citet{caldiroli_irradiation-driven_2021-1}, who (excluding stable atmospheres) find efficiencies in the range 30\%-90\% for planets with $F_{XUV}/\rho_p < 10^4$~erg~cm~s$^{-1}$~g$^{-1}$. We also find lower efficiencies than \citet{vissapragada_upper_2022}, who fit the efficiency parameter for a sample of five planets whose mass-loss rates they constrained using Parker wind models. They report an efficiency of $0.41^{+0.16}_{-0.13}$, but we note that they divided their mass-loss rates by a factor 4 to account for flux averaging over the planet surface, so that their efficiency definition implicitly differs from ours. Although the discrepancies with the latter three works are interesting and potentially worthwhile exploring in more detail, our sample size is currently not large enough to draw definitive conclusions about the value of $\eta$.

\section{Summary} \label{sec:summary}
Understanding the atmospheric mass loss of an exoplanet is crucial for understanding the planet's evolution. Constraining mass-loss rates is commonly done by fitting spectroscopic observations with a set of isothermal Parker wind models, which parametrize the outflow in the mass-loss rate and temperature. This approach often results in a strong degeneracy between these parameters, particularly so for spectrally unresolved observations. We aim to resolve this degeneracy by placing prior constraints on the temperature.

We do this by simulating Parker wind models with the photoionization code \texttt{Cloudy}. We develop an algorithm around the code to include hydrodynamical effects on the temperature structure that are normally not supported. The algorithm works by running \texttt{Cloudy} with a fixed temperature profile and combining the resulting radiative heating rates with advection and expansion terms to propose a new temperature structure based on the heating to cooling ratio. After a few iterations, this yields a converged temperature structure with the hydrodynamic terms included. Our code is relatively fast to execute, and by virtue of using \texttt{Cloudy} includes a highly detailed treatment of the stellar SED and NLTE processes in the upper atmosphere. 

We find different thermal structures for planets with relatively high gravitational potential like HD~209458~b and lower gravity planets like WASP-69~b. The atmospheres of high-gravity planets ($\vert \phi\vert \gtrsim 10^{12.82}$~erg~g$^{-1}$) are characterized by a transition from radiative cooling to expansion work cooling at higher altitudes. Low-gravity giant planets cool by expansion work even at low altitudes, making the temperature at the base of the atmosphere dependent on the assumed outflow velocity. 

We run grids of Parker wind profiles for HD~209458~b, WASP-69~b, WASP-52~b, HAT-P-18~b, HAT-P-11~b and WASP-107~b, which all have metastable helium observations in the literature. For each Parker wind profile, we calculate the difference between the mean temperature in the He 10830~Å line and the isothermal value assumed to create the model. Treating this difference as a quantifier for the self-consistency of each model, we identify a self-consistent parameter space for each planet. We then calculate the transmission spectrum in the He 10830~Å line for every model and fit these to the observed values. Combining the self-consistent parameter spaces and the best-fit Parker wind profiles of the observations give much better constraints on both the temperature and mass-loss rate. In this way, for HD~209458~b, WASP-69~b, WASP-52~b, HAT-P-11~b, HAT-P-18~b and WASP-107 b, we find mass-loss rates of $\dot{M}=10^{9.84^{+0.24}_{-0.27}}$~g~s$^{-1}$, $\dot{M} = 10^{9.91^{+0.19}_{-0.12}}$, $\dot{M} = 10^{11.06^{+0.26}_{-0.25}}$, $\dot{M} = 10^{9.70^{+0.29}_{-0.15}}$, $\dot{M} = 10^{10.33^{+0.29}_{-0.24}}$ and $\dot{M} = 10^{9.96^{+0.23}_{-0.19}}$~g~s$^{-1}$, respectively. In an energy-limited framework, these mass-loss rates would indicate heating efficiencies of 2\%-30\%.

\begin{acknowledgements}
      We thank the anonymous referee for providing valuable comments and suggestions that improved the manuscript. We are grateful to G. J. Ferland and other developers for making \texttt{Cloudy} publicly available, as well as providing support on the help forum\footnote{https://cloudyastrophysics.groups.io}. We appreciate the helpful discussions with J. Kirk, C. Dominik, F. Nail and P. Uttley. We thank SURFsara\footnote{www.surfsara.nl} for the support in using the Lisa Compute Cluster. AO gratefully acknowledges support from the Dutch Research Council NWO Veni grant.
\end{acknowledgements}

\bibliographystyle{aa}
\bibliography{library}

\begin{appendix}
\section{Using the fixed temperature to fit observations} \label{sec:app}
In Sec. \ref{sec:constraints}, we fit a grid of Parker wind models to observations in the He 10830~Å line, resulting in the constraints shown in blue in Fig. \ref{fig:constrain_others}. We used the temperature structures and metastable helium densities as given by the converged \texttt{Cloudy} simulations to calculate the synthetic spectral lines of the models. Here, we investigate how these constraints change if we fit Parker wind models with their assumed fixed temperature $T_0$. This has been the standard approach in literature, and for the observational data we use in this work, such an analysis has been performed by \citealt{lampon_modelling_2020} for HD~209458~b, \citealt{vissapragada_constraints_2020} for WASP-69~b, \citealt{mansfield_detection_2018} for HAT-P-11~b and \citealt{paragas_metastable_2021} for HAT-P-18~b.

\begin{figure*}
   \includegraphics[width=\hsize]{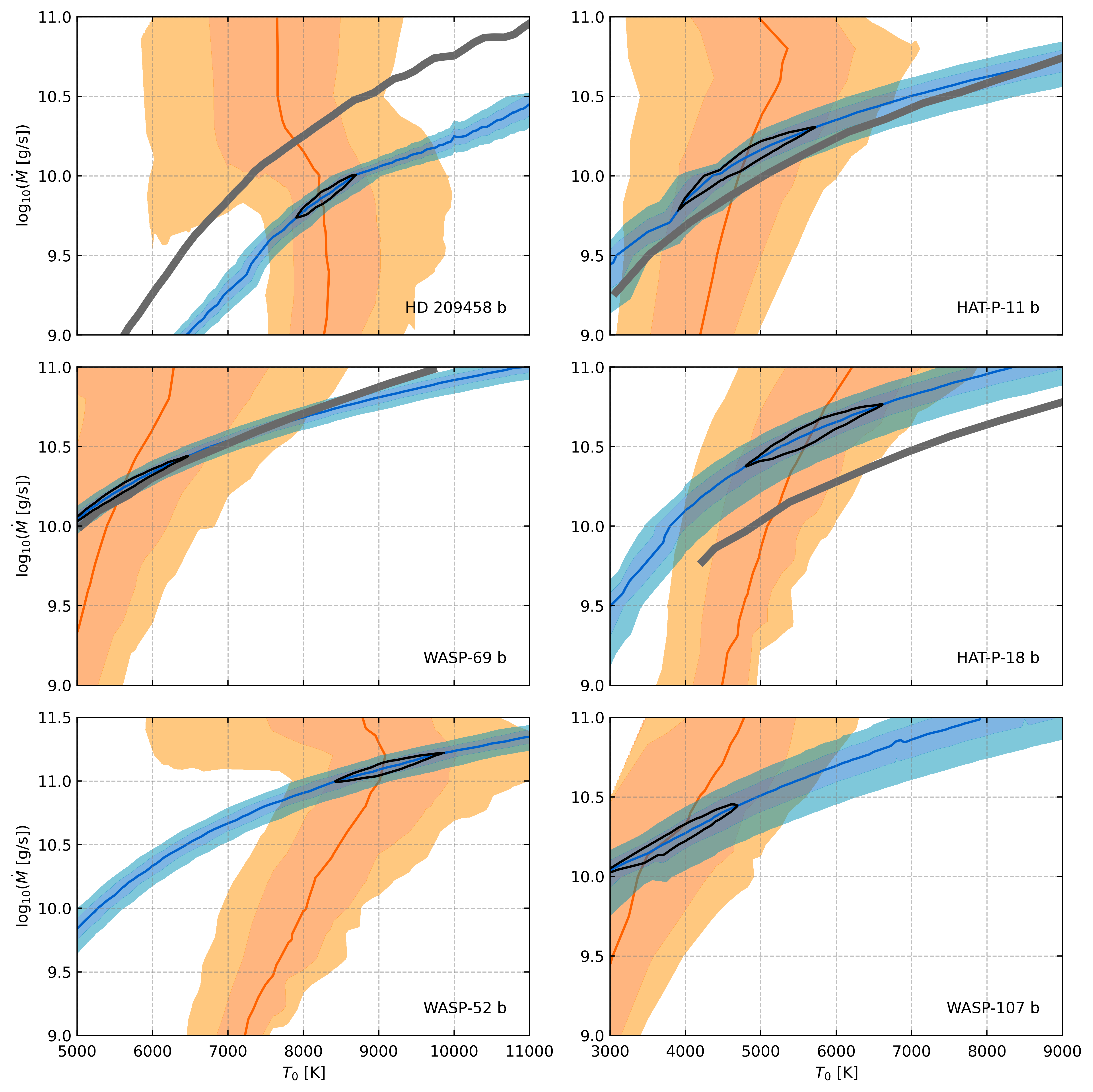}
      \caption{Similar to Fig. \ref{fig:constrain_others}, but using the assumed fixed temperature $T_0$ instead of \texttt{Cloudy}'s temperature $T(r)$ to calculate the synthetic spectral lines. As a consequence, the observational fits shown in blue are different. The thick grey lines indicate the best-fit Parker wind models from similar analyses in the literature: \citealt{lampon_modelling_2020} for HD~209458~b, \citealt{vissapragada_constraints_2020} for WASP-69~b, \citealt{mansfield_detection_2018} for HAT-P-11~b and \citealt{paragas_metastable_2021} for HAT-P-18~b (who used a transit depth of $0.46 \pm 0.12$~\% that is lower than our used value of $0.70 \pm 0.16$~\%). The thickness of the grey lines is arbitrary and does not represent the confidence intervals of those studies.}
         \label{fig:constrain_others_f}
   \end{figure*}

We used \texttt{Cloudy} to simulate a grid of Parker wind models at a constant temperature $T_0$ for each planet. Fig. \ref{fig:constrain_others_f} shows the constraints we find for these simulations. For HD~209458~b and WASP-52~b (the planets with the deepest gravitational potential wells of our sample) the results are very similar to those of Fig. \ref{fig:constrain_others}, but for the other planets we find a stronger degeneracy between the mass-loss rate and temperature. This stems from the fact that the synthetic helium lines are different for simulations with fixed or converged temperature structures, which result in different abundances of helium atoms in the metastable state. We observe that this is particularly the case for the lower gravity planets, likely because they have a qualitatively different energy balance than higher gravity planets (see Sec. \ref{sec:twotypes}) in the region where the helium line forms (typically $r\lesssim 2R_p$).

Comparing the observational constraints of Fig. \ref{fig:constrain_others_f} to Fig. \ref{fig:constrain_others} reveals that simulating the temperature structure of Parker wind models with \texttt{Cloudy} instead of using the assumed temperature already results in better constrained mass-loss rates for lower gravity planets, even when no priors are placed on the self-consistent parameter space. Further comparing our observational constraints of WASP-69~b and HAT-P-11~b of Fig. \ref{fig:constrain_others_f} to the corresponding results in literature (indicated by thick grey lines), we typically find a very similar shape and offsets on the order of $\lesssim$0.2~dex in mass-loss rate. We expect these differences to be mostly due to the use of \texttt{Cloudy}. For HAT-P-18~b, the offset is a bit larger, but also partly stems from the fact that we used the updated transit depth from \citet{vissapragada_upper_2022} that is higher than the value used in \citet{paragas_metastable_2021}. In the case of HD~209458~b, the difference with the constraints from \citet{lampon_modelling_2020} is on the order or $\sim$0.5~dex, which might be due to fitting the EW instead of the resolved line shape.

\end{appendix}

\end{document}